\documentstyle[prd,aps,eqsecnum,preprint,tighten,psfig,floats]{revtex}
\begin{document}
\draft


\preprint{\vbox{\baselineskip=14pt%
   \rightline{FSU-HEP-970527}\break
      \rightline{DESY 97-111}
}}

\title{Charm quark and $D^{\ast \pm}$ cross sections in deeply 
inelastic scattering \\ at HERA}

\author{B. W. Harris}
\address{Physics Department \\ Florida State University \\
Tallahassee, Florida 32306-4250, USA}

\author{J. Smith\footnote{On leave from ITP, SUNY at Stony Brook, Stony 
Brook, NY 11794-3840, USA}}
\address{DESY \\ Notkestrasse 85 \\
D-22603 Hamburg, Germany}
 
\date{May 1997}

\maketitle

\begin{abstract}
A next-to-leading order Monte Carlo program for the calculation of 
heavy quark cross sections in deeply inelastic scattering is described.  
Concentrating on charm quark and $D^{\ast \pm}(2010)$ production 
at HERA, several distributions are presented and their variation 
with respect to charm quark mass, parton distribution set, and 
renormalization/factorization scale is studied.

\end{abstract}


\section{Introduction}
Electromagnetic interactions have long been used to study both 
hadronic structure and strong interaction dynamics.  Examples include 
deeply inelastic lepton-nucleon scattering, hadroproduction of 
lepton pairs, the production of photons with large 
transverse momenta, and various photoproduction processes involving 
scattering of real or very low mass virtual photons from hadrons.  
In particular, heavy quark production in deeply inelastic 
electron-proton scattering (DIS) is calculable in QCD and 
provides information on the gluonic content 
of the proton which is complementary to that obtained in direct 
photon production or structure function scaling violations.  
In addition, it forces one to address such issues as when the 
mass of the heavy quark should be neglected, and how this is 
done consistently.

Activity in the area of neutral current DIS charm quark production 
has increased recently with new data becoming available from 
ZEUS \cite{zeus} and H1 \cite{h1} at HERA.  In particular, 
substantial samples of $D^{\ast \pm}(2010)$ hadrons have been 
obtained. On the theoretical side, much attention has been given to 
the study of heavy quark contributions to the proton structure function.  
They have been 
calculated to next-to-leading order in fully inclusive \cite{LRSvN1}, 
single inclusive \cite{LRSvN2}, and fully differential (exclusive)
\cite{HS} forms as QCD corrections to the photon-gluon fusion process 
(i.e.\ in ${\cal O}( \alpha_s^2) + {\cal O}( \alpha_s^3)$ using three 
flavor parton densities).
Experimentally \cite{zeus,h1,emc} and 
phenomenologically \cite{or,grs}, it is this process 
that is seen to dominate near the threshold region.  
However, well above threshold the heavy quark may 
be considered massless and included in the parton distribution 
function of the proton.  
Various schemes to match these two regions have been proposed 
\cite{var,mrrs,buza}.  Despite all of the attention 
structure functions have received, much less has been done to explore the 
actual heavy quark differential cross sections in DIS which are, in fact, 
much easier to measure experimentally.

The purpose of this work is to present next-to-leading order 
cross sections for charm quark production in the $x$ and $Q^2$
region covered by the HERA collider.  Additionally, 
predictions for heavy hadrons, namely $D^{\ast \pm}(2010)$, are given.
The calculation is implemented in a Monte Carlo style program which 
allows the simultaneous histogramming of many distributions incorporating 
experimental cuts.  It represents an elaboration of the brief results 
already presented by one of us \cite{harris}, and an application of the fully 
differential heavy quark structure functions calculated in \cite{HS}.
No experimental data is shown here because the 
results have already been added to several of the plots in \cite{zeus} 
(see also \cite{harris}).  
An extensive comparison will be given elsewhere.
Herein, the calculation itself is discussed and the variation 
with respect to the theoretical parameters is studied.
Heavy quark correlations have also been calculated for hadroproduction 
\cite{mnr}, photoproduction \cite{fmnr}, and photon-photon collisions 
\cite{kl,ho} allowing for the possiblility of an extensive comparison 
with experimental data.

The calculation was performed using the subtraction method which is based 
on the replacement of divergent (soft or collinear) terms in the squared 
matrix elements by generalized distributions.  The method was first used 
in the context of electron-positron annihilation \cite{sub} and its 
essence is described and compared to the phase-space slicing 
method \cite{phase} in the introduction of a paper by 
Kunszt and Soper \cite{ks}.

The remainder of the paper is as follows.  A review of the subtraction method 
and other aspects of the calculation, including how the hadronization is 
modeled, are given in Sec.\ II.
Numerical results and a discussion of related physics
issues are presented in Sec.\ III.  The conclusions are given in 
Sec.\ IV.

\section{Calculation}
In this section the calculation of the charm quark cross section in 
deeply inelastic scattering is described.  
First, the cross section is written in terms of the charm quark 
contribution to the proton structure functions.  
Then the next-to-leading order QCD corrections to the structure 
functions are reviewed.  
Finally we describe the connection with the production of heavy hadrons 
containing a charm quark.

\subsection{Cross section in terms of structure functions}
The reaction under consideration is charm quark production {\em via} 
neutral-current electron-proton scattering.
\begin{equation}
\label{reaction}
e^{-}(l) + P(p) \rightarrow e^{-}(l') + Q(p_1) + X\,.
\end{equation}
When the momentum transfer squared 
$Q^2=-q^2>0$ ($q=l-l'$) is not too large $Q^2 \ll M_Z^2$, 
the contribution from $Z$ boson exchange is kinematically suppressed 
and the process is dominated by virtual-photon exchange.  
The cross section may then be 
written in terms of structure functions $F^c_2(x,Q^2,m)$ and $F^c_L(x,Q^2,m)$ 
which depend explicitly on the charm quark mass $m_c$ \cite{LRSvN1,schuler} 
as follows:
\begin{equation}
\label{cross}
\frac{d^2\sigma}{dydQ^2} 
= \frac{2\pi\alpha^2}{yQ^4} \left\{ \left[ 1 + (1-y)^2 \right] 
F^c_2(x,Q^2,m_c) - y^2 F^c_L(x,Q^2,m_c) \right\}
\end{equation}
where $x = Q^2 / 2p \cdot q$ and $y = p\cdot q / p \cdot l$ 
are the usual Bjorken scaling variables and $\alpha$ is 
the electromagnetic coupling.  The scaling variables are related to the 
square of the c.m.\ energy of the electron-proton system $S=(l+p)^2$ 
{\em via} $xyS=Q^2$.  The total cross section \cite{schuler} is given by
\begin{equation}
\label{cross2}
\sigma = \int_{4m_c^2/S}^1 dy \int_{m_e^2 y^2/(1-y)}^{yS-4m_c^2} dQ^2
\left( \frac{d^2\sigma}{dydQ^2} \right)
\end{equation}
where $m_e$ is the electron mass.  In deriving 
Eq.\ (\ref{cross}), one integrates over the azimuthal angle between the 
plane containing the incoming and outgoing electrons and the plane containing 
the incoming proton and the outgoing charm quark.

As mentioned in the introduction, this process 
is described near threshold in the framework of perturbative QCD by 
flavour creation through the virtual-photon-gluon fusion process
\begin{equation}
\label{fusion}
\gamma^*(q) + g(k_1) \rightarrow Q(p_1) + \bar{Q}(p_2).
\end{equation}
The structure functions follow \cite{LRSvN1,HS} from the longitudinal 
$\sigma_L$ and 
transverse $\sigma_T$ cross sections for this reaction {\em via} 
$F^c_2=(Q^2 / 4 \pi^2 \alpha) \, ( \sigma_L + \sigma_T )$ 
and $F^c_L=(Q^2 / 4 \pi^2 \alpha) \, \sigma_L$.  Thus, QCD corrections 
to the reaction (\ref{reaction}) correspond to QCD corrections 
to (\ref{fusion}) to which we now turn.  The supersript $c$ will 
henceforth be dropped to simplify notation.

\subsection{QCD corrections to the heavy quark structure functions}
Within the context of perturbative QCD, structure functions are expressed 
as a product of the running coupling, the parton densities, and 
the hard scattering cross sections.
The result is a physical quantity, but the individual terms can
be defined in a convenient scheme which moves terms from one factor
to another.  All schemes should give the same result for the product, 
up to terms of higher order.  

A next-to-leading order calculation of the heavy quark contributions to the 
proton structure functions requires the one-loop virtual corrections 
to (\ref{fusion}). 
For this set of diagrams, the renormalization was carried out so that 
divergences coming from the
light quarks were subtracted in the standard $\overline{\rm MS}$ scheme,
while the divergences coming from heavy quark loops were subtracted at 
zero external momenta.  This is the scheme originally proposed in  
\cite{cwz} in which the mass $m$ only appears in the hard scattering 
cross sections.  As a consequence, below the subtraction scale, 
only the number of (massless) light 
flavours appears in the running coupling 
and in the splitting functions used in the parton evolution equations. 
At the subtraction scale $\mu=m$ there are matching 
conditions for both the running coupling and the 
light flavour densities \cite{cwz}.  Therefore, to order $\alpha_s$, 
there is no charm density 
at the scale $\mu = m$ \cite{cwz}, \cite{nde}.

Consequently, one should use only parton distribution sets that 
were derived from data using the same renormalization scheme.  
Examples of such densities are GRV94\cite{grv94} and CTEQ4F3\cite{cteq4f3}.
At larger scales there is a charm density proportional to $\alpha_s
\ln (\mu^2/m_c^2)$, which grows at the expense of a reduction in the 
gluon density.  One of the interesting problems in
the analysis of charm quark contributions to deeply inelastic scattering 
is to understand the transition region from the photon-gluon predictions 
based on three light flavours
to a charm density picture with four light flavours.
The matching conditions become more complicated as one goes to 
higher order.  The corresponding two-loop matching conditions for 
the flavour densities have been calculated in \cite{buza} 
wherein they were found to have finite terms at $\mu = m$.

In addition to the virtual corrections described above, there is 
also a contribution from the gluon-bremsstrahlung process
\begin{equation}
\label{brem}
\gamma^*(q) + g(k_1) \rightarrow Q(p_1) + \bar{Q}(p_2) + g(k_2)
\end{equation}
and new production mechanisms not present at leading order, 
which are given by 
\begin{eqnarray}
\label{quark}
\gamma^*(q) + q(k_1) & \rightarrow & Q(p_1) + \bar{Q}(p_2) 
+ q(k_2) \nonumber \\
\gamma^*(q) + \bar{q}(k_1) & \rightarrow & Q(p_1) + \bar{Q}(p_2) 
+ \bar{q}(k_2)
\end{eqnarray}
where $(\bar{q})q$ is a massless (anti-)quark.
The contribution to the structure functions resulting 
from these processes 
have been calculated in a fully differential \cite{HS} 
form and are suitable for use in constructing a
Monte Carlo style program because one has full access to the final 
state partonic four vectors.

Briefly, the computation in \cite{HS} was carried out using the 
subtraction method which is
based on the replacement of divergent (soft or collinear) terms in
the squared matrix elements by generalized plus distributions.
This allows one to isolate the soft and collinear poles within the
framework of dimensional regularization without calculating all the
phase space integrals in a space-time dimension $n \ne 4$ as is 
required in a traditional single particle inclusive computation.
In this method the expressions for the squared matrix elements in the
collinear limit appear in a factorized form, where poles in $n-4$ multiply
splitting functions and lower order squared matrix elements.
The cancellation of collinear singularities is then performed using
mass factorization.  The expressions for the squared matrix
elements in the soft limit also appear in a factorized form where poles in
$n-4$ multiply lower order squared matrix elements.  The cancellation of soft
singularities takes place upon adding the contributions from the renormalized
virtual diagrams.  Since the final result is in four-dimensional space time,
one can compute all relevant phase space integrations using standard Monte
Carlo integration techniques.  

The resultant (differential) structure functions may be written in the form 
\begin{eqnarray}
\label{fhad}
F_k(x,Q^2,m) &=& \frac{Q^2 \alpha_s(\mu^2)}{4\pi^2 m^2}
\int_{\xi_{\rm min}}^1 \frac{d\xi}{\xi} \left\{ \left[ c^{(0)}_{k,g} +
4 \pi \alpha_s(\mu^2) \left( c^{(1)}_{k,g}
+ \bar c^{(1)}_{k,g} \ln \frac{\mu^2}{m^2} \right)
\right] e_H^2 f_{g/P}(\xi,\mu^2) \right. \nonumber \\
&+& \left. 4 \pi \alpha_s(\mu^2) \sum_{i=q,\bar q} f_{i/P}(\xi,\mu^2)
\left[ e_H^2 \left( c^{(1)}_{k,i} + \bar c^{(1)}_{k,i} \ln \frac{\mu^2}{m^2}
\right) + e^2_i \, d^{(1)}_{k,i} + e_i\, e_H \, o^{(1)}_{k,i} \,
\right] \right\} \nonumber \\ &&
\end{eqnarray}
$k = 2,L$.
The lower boundary on the integration is $\xi_{\rm min} = x(4m^2+Q^2)/Q^2$.
The parton momentum distributions in the proton are denoted by
$f_{i/P}(\xi,\mu^2)$.  The mass factorization scale $\mu_f$ has been 
set equal to the renormalization scale $\mu_r$ and is denoted by $\mu$.
All charges are in units of $e$. 
Finally, $c^{(0)}_{k,i}$, $c^{(1)}_{k,i}$, 
$\overline c^{(1)}_{k,i}\,, (i = g, q, \bar q)$,  and 
$d^{(1)}_{k,i} $, $o^{(1)}_{k,i} \,,(i= q, \bar q)$ are scale 
independent parton coefficient functions.  
They are functions of $\xi$, $Q^2$, and $m$.
In Eq.\ (\ref{fhad}) the coefficient functions 
are distinguished by their origin. The $c$-coefficent functions
originate from processes involving the virtual photon-heavy quark coupling, 
while the $d$-coefficient functions arise from processes involving the
virtual photon-light quark coupling and the $o$-coefficent functions are
from the interference between these processes. 

In addition to the calculation of \cite{HS}, the functions 
$c^{(1)}_{k,i}$, $\overline c^{(1)}_{k,i}$, and 
$d^{(1)}_{k,i}$ were calculated in inclusive form  
in \cite{LRSvN1},\cite{LRSvN2} as two-dimensional integrals and
computed numerically.  
In \cite{steve} they were numerically tabulated in 
grids, with a fast interpolation routine, so that the computation
of Eq.\ (\ref{fhad}) could be included in a global fit, if desired.
Finally, in \cite{buzaa} exact analytic formulae
were given for the $d^{(1)}_{k,i}$ together with analytic formulae
for all the coefficient functions
$c^{(0)}_{k,i}$, $c^{(1)}_{k,i}$, 
$\overline c^{(1)}_{k,i}\,, (i = g, q, \bar q)$,  
and $d^{(1)}_{k,i}\,, (i= q, \bar q)$ in the limit $Q^2 >> m^2$.  
The latter results are necessary to consider a variable flavour scheme 
in which the coefficient functions are incorporated into redefined 
light parton densities including a charm density \cite{buza}.

Naturally, properties of the structure functions give insight into the 
behavior of the cross section.
Therefore the more salient features of the next-to-leading order 
structure functions will now be summarized.  
The interested reader can find additional details in the original papers 
\cite{LRSvN1,LRSvN2,HS} and, more so, in the recent 
phenomenological analyses \cite{grs,vogt,work1,work2}.  
For moderate $Q^2 \sim 10\,{\rm GeV}^2$ one finds that the charm quark 
contribution at small $x \sim 10^{-4}$ is 
approximately $25\%$ of the total structure function (defined as light parton 
plus heavy quark contributions).  In contrast, the contribution from bottom 
quarks is only a few percent due to charge and phase space suppression.  
Thus, the charm quark contribution must be retained, but the bottom quark 
contribution may be neglected in the following.
The gluon initiated contributions (\ref{fusion}) and (\ref{brem}) 
comprise most of the structure function.  The quark initiated processes 
(\ref{quark}) give only a few percent contribution at small $x$ 
for reasonable scale choices.  
Results for the charm quark contribution to the proton structure function 
$F_2^c(x,Q^2,m_c=1.5\, {\rm GeV})$ as a 
function of $x$ are shown in FIG.\ 1 for $\mu=\sqrt{Q^2+4m_c^2}$ using 
the GRV94 \cite{grv94} and CTEQ4F3 \cite{cteq4f3} parton distribution sets.
\begin{figure}
\centerline{\hbox{\psfig{figure=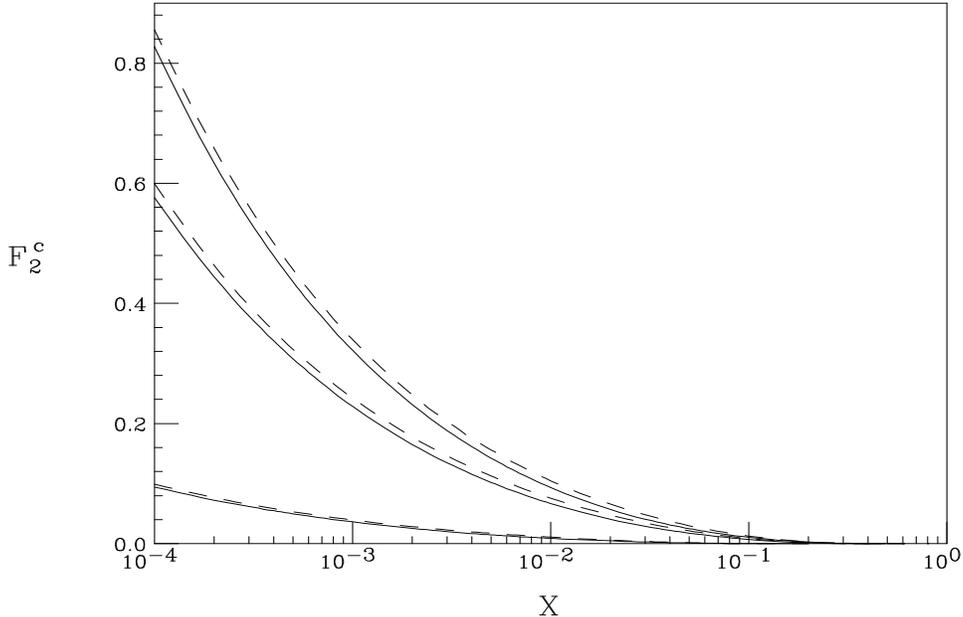,width=5.0in,height=3.5in}}}
\caption{$F_2^c(x,Q^2,m_c=1.5\,{\rm GeV})$ as a function of $x$ for GRV94 
(solid lines) and CTEQ4F3 (dash lines) parton distribution sets for 
$Q^2=3\, {\rm GeV}^2\,$(bottom), $25\, {\rm GeV}^2\,$(middle), and $50\, 
{\rm GeV}^2\,$(top).}
\end{figure}
As mentioned above, 
these sets were chosen because they have $n_f=3$ in the evolution and 
are therefore the most consistent sets to use with the NLO calculation.  
The curves show a marked rise in the structure function at small $x$ 
due primarily to the rapidly rising gluon distribution.  
The renormalization/factorization scale dependence is rather flat, 
especially at small $x$ where the structure function is largest.  
This is demonstrated in FIG.\ 2 for various $x$ and $Q^2$ values.  
\begin{figure}
\centerline{\hbox{\psfig{figure=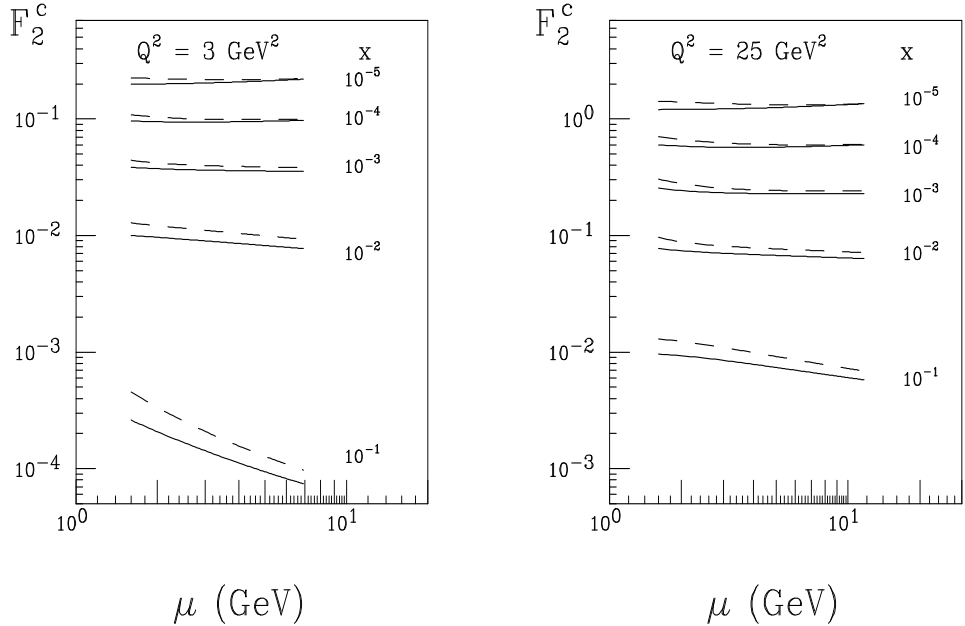,width=5.0in,height=3.5in}}}
\caption{$F_2^c(x,Q^2,m_c=1.5\,{\rm GeV})$ as a function of the scale $\mu$ 
with $m_c \leq \mu \leq 2 \protect\sqrt{Q^2+4m_c^2}$ for various $x$ and 
$Q^2$ values for the GRV94 (solid lines) and CTEQ4F3 (dash lines) parton 
distribution sets.}
\end{figure}
By far the largest uncertainty in the calculation of the structure 
functions is the value of the heavy quark mass.  For charm production, 
for example, a $\pm 10\%$ variation of the mass about the central value of 
$m_c=1.5\,{\rm GeV}$ gives a $\pm 20\%$ variation in the structure function 
for moderate $Q^2$.

\subsection{Fragmentation into heavy hadrons}
Experimentally, it is heavy hadrons or their decay productions 
that are observed.  Of interest for HERA is $D^{\ast \pm}(2010)$ 
production.  Herein, the Peterson {\it et al}.\ form of the fragmentation 
function \cite{pete} is used to model the nonperturbative hadronization 
process.

The cross section for $D^{\ast}$ production is obtained by convoluting 
the charm quark cross section (\ref{cross2}) with the fragmentation function 
\begin{equation}
D(z)=\frac{N}{z[1-1/z-\epsilon/(1-z)]^2}
\end{equation}
where $N$ is fixed such that $D(z)$ is normalized to unity.  The 
normalization of the cross section is then fixed by the charm quark 
fragmentation probability $P(c \rightarrow D^*)=0.26$ \cite{norm}.
The parameter $\epsilon$ may be extracted from data \cite{eps0} and used as 
input.  However, in light of recent work on the subject \cite{eps1,eps2}, 
the specific value that should be used in this particular application is 
not obvious.  The choice of the best value is left as the subject for 
future study.  Below $\epsilon$ is treated as a free parameter, and the 
effect of its variation on the cross section will be examined and 
considered as part of the uncertainty due to hadronization.

Other sources of uncertainty include such technical details as how 
exactly the four vector of the $D^{\ast}$ is formed.  One may scale 
the entire four vector by $z$, but then the hadron mass is $zm_c$.  Another 
possibility is to scale the three vector by $z$ and fix the energy 
component such that that the mass is $m(D^{\ast})=2.01\, {\rm GeV}$.  
The latter is used here.

Evolution of the fragmentation function, which one expects to become 
important when $p_t \gg m_c$, is not included because the region of 
interest is $p_t \sim m_c$.  Indeed, recent 
calculations of charm photoproduction at HERA \cite{evol} have shown 
that this effect becomes sizeable only for $p_t > 20\, {\rm GeV}$.  

\section{Results}
Using the results of the method described in the previous section, a computer 
program has been constructed to calculate charm quark and/or $D^*$ cross 
sections in deeply inelastic scattering 
\footnote{Interested readers should contact\, {\tt harris@hep.fsu.edu}\, for 
a copy of the computer code.}.
The program uses Monte Carlo integration so it is 
possible to study a variety of distributions and 
implement experimental cuts, provided they are defined in 
terms of partonic variables or the optional fragmentation into heavy hadrons 
is used.  Results are presented in the HERA lab frame with 
positive rapidity in the proton direction.  The proton and electron beam 
energies are taken to be $820\, {\rm GeV}$ and $27.6\, {\rm GeV}$, 
respectively.  
There are several necessary cross checks that should be performed before 
discussing the general properties of the complete next-to-leading order 
calculation.

\subsection{Comparison with existing results}
\begin{figure}
\centerline{\hbox{\psfig{figure=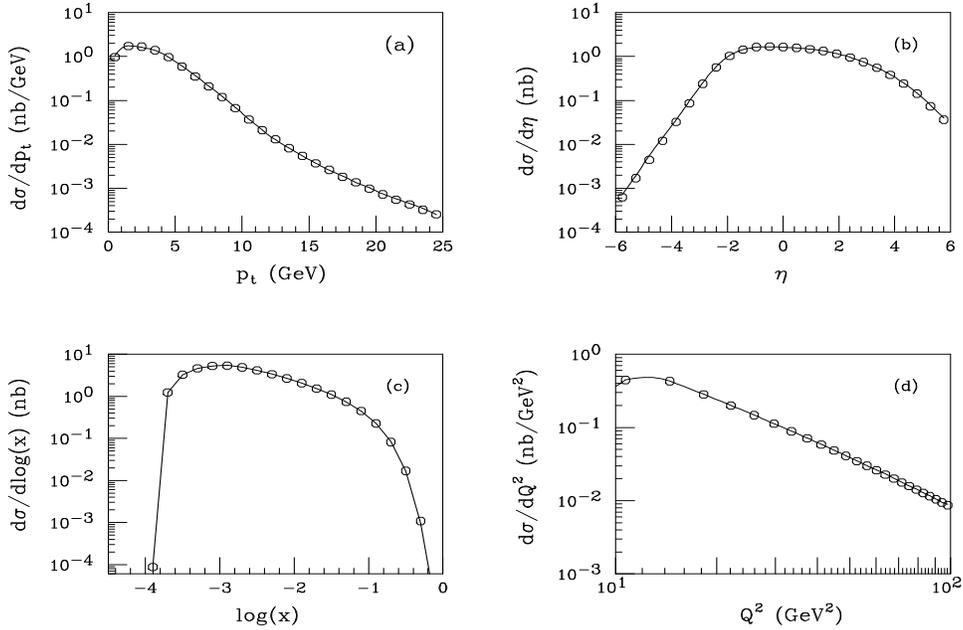,width=5.0in,height=3.5in}}}
\caption{Comparison of leading order results (open circles) with 
{\sc AROMA} \protect\cite{aroma} (solid lines) for charm production 
in the kinematic region $10<Q^2<100\, {\rm GeV}^2$ and $0<y<0.7$ at HERA.}
\end{figure}
A comparison with the leading order event generator {\sc AROMA} 
\cite{aroma} provides a check on the kinematics and leading order 
matrix elements through the shape and overall normalization of the 
distributions.  For this purpose, both codes were run with the 
CTEQ2L \cite{cteq123} proton parton distributions, the default set for 
{\sc AROMA}. This set has $\Lambda^{(4)}=190\,{\rm MeV}$ which 
was used, along with $n_f=4$, in the one-loop strong coupling $\alpha_s$.
The renormalization and factorization scales were both set to 
$\sqrt{\hat{s}}$.  In the {\sc AROMA} calling program all fragmentation 
and showering was turned off, and only the contribution from virtual photon 
exchange was retained.  Shown in FIG.\ 3 are the results for charm quark 
production, assuming $m_c=1.5\, {\rm GeV}$, in the kinematic region 
$10<Q^2<100\, {\rm GeV}^2$ and $0<y<0.7$.   The distributions shown 
are for the transverse momentum $p_t$ and pseudo-rapidity $\eta$ of the 
charm quark, both in the HERA lab frame, along with the usual deep 
inelastic scattering variables $x$ and $Q^2$.  These will be the 
canonical set of observables used in the rest of the paper. 
The shapes of the curves are virtually identical over several orders 
of magnitude and 
the area under the curves is the same to better than $1\%$.

Another check is to reproduce the 
total heavy quark cross section as previously calculated in next-to-leading 
order \cite{steve}.  Both computer programs were run with the 
CTEQ3M \cite{cteq123} proton parton distributions.  This set has 
$\Lambda^{(4)}=239\,{\rm MeV}$ which was used, along with $n_f=4$, 
in the two-loop strong coupling $\alpha_s$.  The renormalization and 
factorization scales were both set to $\sqrt{Q^2+4m_c^2}$. The results 
for charm quark production, again assuming $m_c=1.5\, {\rm GeV}$, in the 
kinematic region $10<Q^2<100\, {\rm GeV}^2$ and $0.01<y<0.7$ are identical 
for both programs to better than three significant figures.  The numerical 
values for the cross section per channel are 7.48 nb for 
${\cal O}(\alpha_s)$ photon-gluon, 2.68 nb for ${\cal O}(\alpha_s^2)$ 
photon-gluon, and $-0.41$ nb for the sum of photon-quark and photon-antiquark.
Having made these checks, the general properties of the 
full next-to-leading order cross section may now be considered.

\subsection{Properties of the charm quark cross section}
\begin{figure}
\centerline{\hbox{\psfig{figure=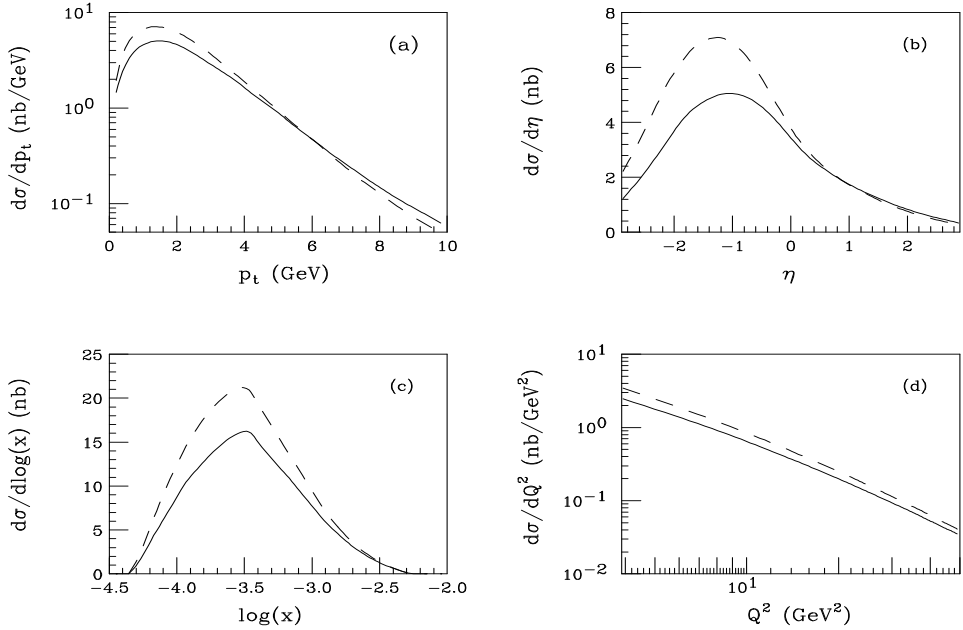,width=5.0in,height=3.5in}}}
\caption{Leading (dash) and full next-to-leading (solid) order differential 
cross sections for charm quark production at $\protect\sqrt{S}=301\, 
{\rm GeV}$ 
at HERA using the GRV94 (LO and HO, respectively) parton distribution set at 
$\mu=\protect\sqrt{Q^2+4m_c^2}$ with $m_c=1.5\, {\rm GeV}$.}
\end{figure}

In this section the dependence of the charm quark cross sections will be 
studied as a function of parton distribution set, charm quark mass, and 
renormalization/factorization scale.  All results are for 
the kinematic range $3<Q^2<50\, {\rm GeV}^2$ and $0.1<y<0.7$.

The CTEQ4F3 \cite{cteq4} and GRV94 HO \cite{grv94} proton-parton distribution 
sets were used in the remainder of the paper, as noted.
For leading order results, the GRV94 LO \cite{grv94} set was used.
The (one-) two-loop version of the strong coupling $\alpha_s$ was
used with matching across quark thresholds for the (LO) NLO results.
The value of $\Lambda_{\rm QCD}$ was taken from the proton-parton 
distribution set.  The renormalization and factorization scales have 
been set equal to $\mu$.

The leading (dash) and next-to-leading (solid) order differential 
cross sections for charm quark production using the GRV94 
(LO and HO, respectively) parton distribution set at 
$\mu=\protect\sqrt{Q^2+4m_c^2}$ with $m_c=1.5\, {\rm GeV}$ are 
shown in FIG.\ 4.  The shape of the NLO transverse momentum 
distribution is similar to the LO one, but somewhat flatter.  The 
pseudo-rapidity distribution shows the radiative corrections 
are concentrated in the negative rapidity direction, tending 
to pull the maximum back towards the central region.  
The Bjorken $x$ distribution receives corrections near its 
maximum with near zero correction at the tails.  The $Q^2$ 
distribution receives a fairly uniform shift in normalization.  
Save at high $p_t$, the NLO predictions lie below the LO 
ones.  This a property of the GRV parton distribution set.  For 
the CTEQ4F3 set the opposite behavior is observed.  This a 
reflection of the difference between the leading order gluon 
distribution functions and the corresponding $\Lambda_{\rm QCD}$ of 
the two sets.

\begin{figure}
\centerline{\hbox{\psfig{figure=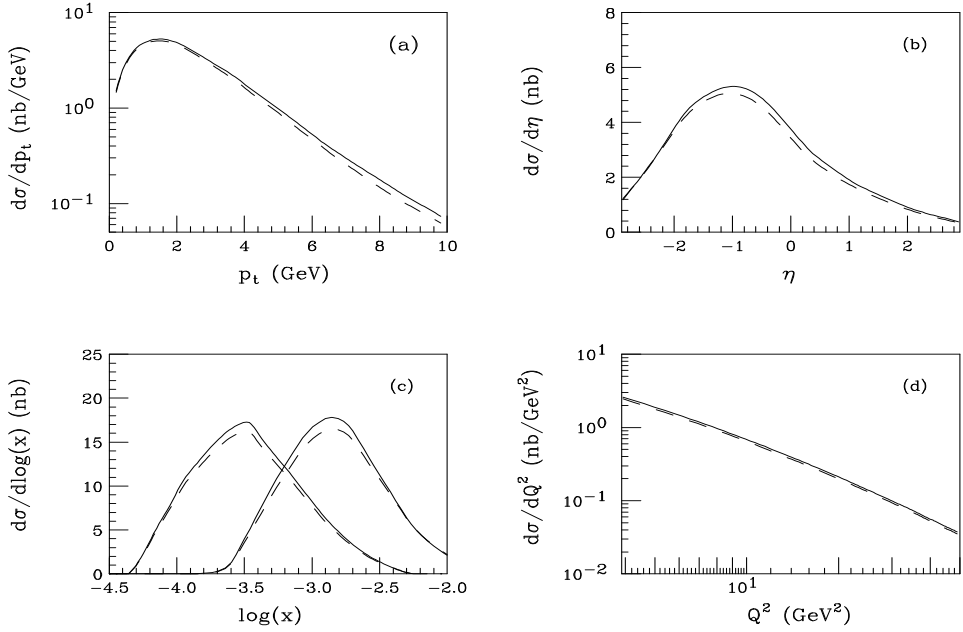,width=5.0in,height=3.5in}}}
\caption{Next-to-leading order differential cross sections for 
charm quark production at $\protect\sqrt{S}=301\, {\rm GeV}$ at HERA 
using the GRV94 (dash) and CTEQ4F3 (solid) parton distribution sets at 
$\mu=\protect\sqrt{Q^2+4m_c^2}$ with $m_c=1.5\, {\rm GeV}$.  Also shown for 
comparison in part $(c)$ is $d\sigma/d\log(\xi)$ vs.\ $\log(\xi)$ (right pair 
of curves).}
\end{figure}

One may ask how sensitive are the full next-to-leading order 
results to modern parton distribution sets.  
The answer is immediate from FIG.\ 5 which shows the 
next-to-leading order differential cross sections for 
charm quark production using the GRV94 (dash) and CTEQ4F3 (solid) parton 
distribution sets at $\mu=\protect\sqrt{Q^2+4m_c^2}$ with $m_c=1.5\, 
{\rm GeV}$.  From Eq.\ (\ref{fhad}) the parton distributions are 
probed at $\xi$ which is typically one order of magnitude larger the 
$x$.  This is illustrated in FIG.\ 5c where a plot 
of $d\sigma/d\log(\xi)$ vs.\ $\log(\xi)$ (right set of curves) 
is superimposed on the plot of $d\sigma/d\log(x)$ vs.\ $\log(x)$ 
(left set of curves).  The difference between the curves produced using 
the two parton distribution sets is approximately $10\%$ at $\xi = 10^{-2.7}$.
Thus, the predictions are less dependent on the parton 
density sets in NLO.

\begin{figure}
\centerline{\hbox{\psfig{figure=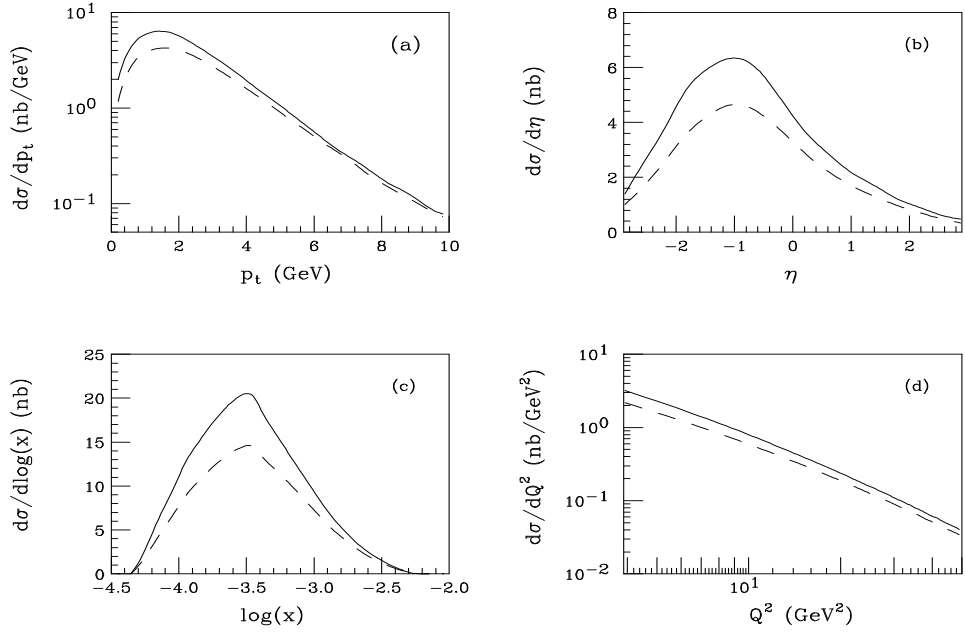,width=5.0in,height=3.5in}}}
\caption{Next-to-leading order differential cross sections for 
charm quark production at $\protect\sqrt{S}=301\, {\rm GeV}$ at HERA 
using the CTEQ4F3 parton distribution set at $\mu=\protect\sqrt{Q^2+4m_c^2}$ 
with $m_c=1.35\, {\rm GeV}$ (solid) and $m_c=1.65\, {\rm GeV}$ (dash).}
\end{figure}

\begin{figure}
\centerline{\hbox{\psfig{figure=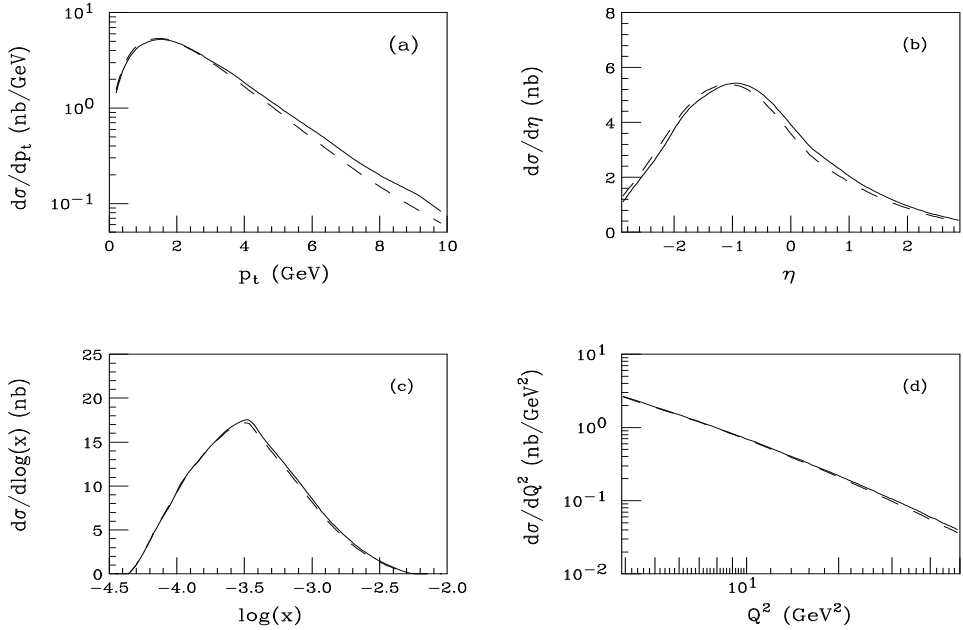,width=5.0in,height=3.5in}}}
\caption{Next-to-leading order differential cross sections for 
charm quark production at $\protect\sqrt{S}=301\, {\rm GeV}$ at HERA 
using the CTEQ4F3 parton distribution set at $\mu=2m_c$ (solid) and 
$\mu=2\protect\sqrt{Q^2+4m_c^2}$ (dash) with $m_c=1.5\, {\rm GeV}$.}
\end{figure}

The largest uncertainty in the structure function calculation is that 
of the charm quark mass.  The same is true for the cross section 
as shown in FIG.\ 6 for the next-to-leading order differential cross 
sections for charm quark production using the CTEQ4F3 parton distribution 
set at $\mu=\sqrt{Q^2+4m_c^2}$ 
with $m_c=1.35\, {\rm GeV}$ (solid) and $m_c=1.65\, {\rm GeV}$ (dash).  
Mass effects are smaller at the larger transverse mass because they are 
suppressed by powers of $m_c/p_t$ in the matrix elements.  
However, as mentioned above, if the 
range is extended much further, large logarithms of the form 
$\ln(p_t^2/m_c^2)$ appear in the cross section and should be resummed.

Finally, the scale dependence is shown in FIG.\ 7.  The next-to-leading 
order differential cross sections are for the CTEQ4F3 parton distribution 
set at $\mu=2m_c$ (solid) and $\mu=2\sqrt{Q^2+4m_c^2}$ (dash) with 
$m_c=1.5\, {\rm GeV}$.  The curves show very little scale dependence.  
This can be anticipated from the results shown in FIG.\ 2 and the 
distribution in Bjorken $x$ shown in FIG.\ 7c.  
The latter shows the cross section is dominated by 
$x \sim 10^{-3.5}=3.2 \times 10^{-4}$ while the former shows that, 
independent of $Q^2$, the structure function is very flat in this particular 
$x$ region.  Therefore, the cross section tends to be fairly insensitive 
to the choice of scale.  
Other kinematic regions show increased scale dependence.  
This serves as a reminder that care must be taken in interpreting 
the results of varying the renormalization/factorization scale to estimate 
the size of the theoretical error.

\subsection{Predictions for $D^*$ production}

In this last section the fragmentation is turned on and 
predictions for $D^{\ast \pm}$ production at HERA are given.
The power of the subtraction method becomes manifest because 
experimental cuts can easily be implemented.  Cuts 
similar to those preferred by ZEUS \cite{zeus} and H1 \cite{h1} are used.  
Namely, $2<Q^2<100\, {\rm GeV}^2$, 
$0.05<y<0.7$, $p_t^{D^{\ast}} > 1.5\, {\rm GeV}$, and 
$| \eta^{D^{\ast}} | < 1.5$.  No distinction is made 
between $D^{\ast +}$ and $D^{\ast -}$.  The results shown in FIG.\ 8 
use the GRV94 parton distribution set at 
$\mu=\sqrt{Q^2+4m_c^2}$ with $m_c=1.5\, {\rm GeV}$ and 
$\epsilon=0.03$ (dot), $\epsilon=0.06$ (solid), $\epsilon=0.09$ (dash).  
The variation in the area under the curves is roughly half that 
from the mass uncertainty.  The shape changes are very mild.  
Using the CTEQ4F3 parton distribution set instead would give slightly 
larger results.

\begin{figure}
\centerline{\hbox{\psfig{figure=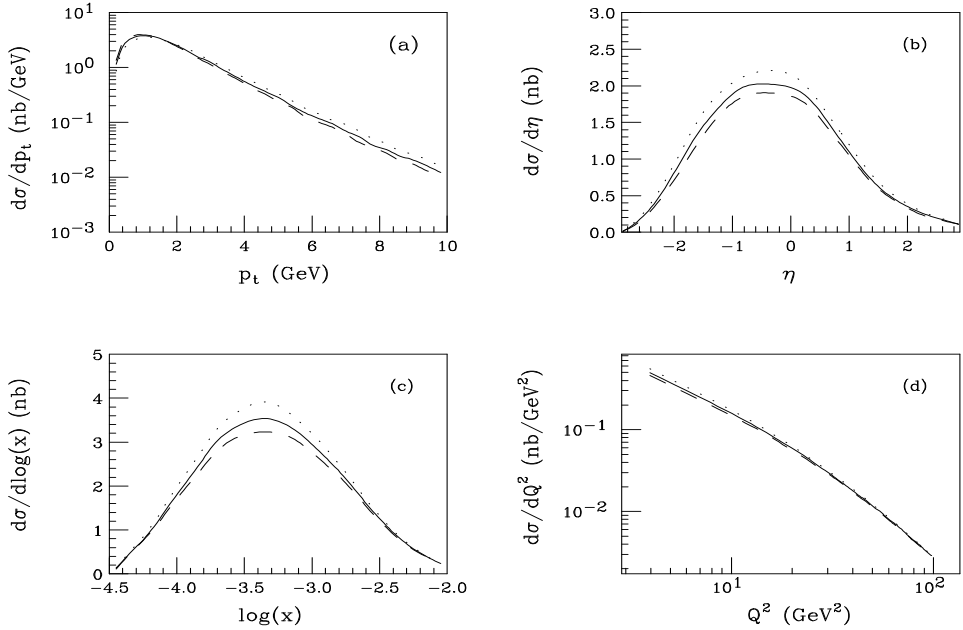,width=5.0in,height=3.5in}}}
\caption{Next-to-leading order differential cross sections for 
$D^*$ production at $\protect\sqrt{S}=301\, {\rm GeV}$ at HERA in 
the kinematic region $0.05<y<0.7$, $2<Q^2<100\, {\rm GeV}^2$, 
$1.5 < P_t^{D^*} < 10\, {\rm GeV}$, and 
$| \eta^{D^*} | < 1.5$ using the GRV94 parton distribution set at 
$\mu=\protect\sqrt{Q^2+4m_c^2}$ with $m_c=1.5\, {\rm GeV}$ and 
$\epsilon=0.03$ (dot), $\epsilon=0.06$ (solid), $\epsilon=0.09$ (dash).}
\end{figure}

\section{Conclusion}
The calculation of next-to-leading order corrections has allowed more reliable 
predictions of heavy quark differential distributions in 
deeply inelastic scattering.  In addition, with the 
calculational formalism used here, it was possible to add 
Peterson fragmentation thus giving predictions for $D^{\ast \pm}(2010)$ 
production at HERA.

At leading order the results were cross checked against 
{\sc AROMA} and found to give complete agreement.  
When the program is run in fully inclusive mode it reproduces 
existing results for the charm quark cross section at 
next-to-leading order.

The radiative corrections to the lowest order photon-gluon 
fusion process are important as they change both 
the shape and normalization of the transverse momentum, 
pseudo-rapidity, and $x$ distributions.  The $Q^2$ distributions 
only receive a shift in normalization.  
In the kinematic region studied, the cross section 
is very stable with respect to variations in 
the renormalization/factorization scale because 
the cross section is dominated by an $x$ region where the 
scale dependence of the underlying structure 
functions is nearly flat.  The scale dependence of the hard scattering 
cross sections is well compensated by that of the parton 
densities and $\alpha_s$.

The cross section is dominated by the rapidly growing gluon 
distribution at small $x$, but distinguishing between modern parton 
distribution sets using this process will be difficult as 
demonstrated by the fact that they give nearly identical 
results for a variety of observables.  
This is compounded by relatively large uncertainties
from the quark mass and hadronization effects.
At present a comparison with experimental data can 
offer a confirmation of the gluon distribution at small $x$. 
Examining a variety of $x$ and $Q^2$ bins will 
shed light on the transition region between massive and massless 
charm quark descriptions.

\acknowledgements

This work was support by contracts DE-FG02-97ER41022 and NSF 93-09888.  
We thank J.F. Owens for comments, J.P. Fern\'{a}ndez 
for discussions concerning the ZEUS analysis, and F. Sefkow and K. Daum
for discussions concerning the H1 analysis.
J. Smith would like to thank the Alexander von Humbolt
Foundation for an award allowing him to spend his sabbatical leave
at DESY.

%
%

\def\Journal#1#2#3#4{{#1} {\bf #2}, #3 (#4)}
\def\AP{\em Annals of Physics}
\def\CPC{\em Comput. Phys. Commun.}
\def\NCA{\em Nuovo Cimento}
\def\NIM{\em Nucl. Instrum. Methods}
\def\NIMA{{\em Nucl. Instrum. Methods} A}
\def\NPB{{\em Nucl. Phys.} B}
\def\PLB{{\em Phys. Lett.}  B}
\def\PRL{\em Phys. Rev. Lett.}
\def\PRD{{\em Phys. Rev.} D}
\def\PR{\em Phys. Rev.}
\def\ZPC{{\em Z. Phys.} C}
\def\ZP{\em Z. Phys.}

\end{document}